\documentstyle[aps,prl,multicol]{revtex}
\input{epsf.tex}
\begin{document}
\draft
\title{A simple theory for high $\Delta / T_{c}$  ratio in d-wave 
superconductors}
\author{R. Combescot and X. Leyronas}
\address{Laboratoire de Physique Statistique, Ecole Normale 
Sup\'erieure*,
24 rue Lhomond, 75231 Paris Cedex 05, France}
\date{Received \today}
\maketitle

\begin{abstract}
We investigate a simple explanation for the high maximum gap to 
$T_{c}$ ratio found experimentally in high $T_{c}$ compounds. We 
ascribe this observation to the lowering of $T_{c}$ by boson scattering 
of electrons between parts of the Fermi surface with opposite sign for 
the order parameter. We study the simplest possible model within this 
picture. Our quantitative results show that we can account for 
experiment for a rather small value of the coupling constant, all the 
other ingredients of our model being already known to exist in these 
compounds. A striking implication of this theory is the fairly high value 
of the critical temperature in the absence of boson scattering.
\end{abstract}
\begin{multicols}{2}
A puzzling feature of high $T_{c}$ cuprate superconductors is the 
fairly high value of the maximum $ \Delta _{0} $ of the 
superconducting gap compared to the critical temperature. Indeed it 
seems to range from 3 to 4 in most experiments, performed mainly on 
YBCO and on BSCCO \cite{rf} . This is to be compared with standard 
BCS value 1.76 . Since it is widely believed that these compounds are 
unconventional, with in particular changes of sign for the order 
parameter, it would seem that this is not much of a problem. However 
this $\Delta_{0} / T_{c}$ ratio is surprisingly very stable within all the 
generalizations of BCS theory which have been put forward for these 
compounds. Van Hove singularities and more generally any varying 
density of states raise it at most up to 2 , any reasonable anisotropy 
\cite{rc} gives a result not so much beyond the d-wave value 2.14  and 
it requires strong coupling effects incompatible with experiments to 
push it in the experimental range. All these explanations are far from 
explaining the typical increase by a factor 2 compared to the BCS value, 
and one may wonder if a more complicated theoretical framework is not 
necessary in order to account for this ratio. 

We show in this paper that this is not the case and that the data can be 
explained quantitatively to a large extent by a simple theoretical model in 
the standard framework of mean-field theory. Actually, except for the 
rather moderate value of the coupling constant we require, all the 
physical ingredients of our model are known to be present in these 
compounds, which makes our explanation a very natural one. However 
we do not apply this claim to the very high values of $\Delta_{0} / 
T_{c}$ found recently \cite{ydw,miy} in underdoped BSCCO. We 
believe that, even if they are clearly related to superconducting 
properties  \cite{miy1}, some additional physics, specific of this regime 
and this very anisotropic compound, is required to account for such 
very high results. Our focus is on the more standard values found 
elsewhere, in other compounds (in particular YBCO) and in optimally 
and overdoped BSCCO.

The basic idea of our model is the following. As is well known, when a 
superconductor has an order parameter which changes sign over the 
Fermi surface, superconductivity tends to be destroyed by anything, 
like impurities, which scatters electrons between parts of the Fermi 
surface with opposite signs. If these scattering sources are present at 
$T_{c}$ but not at $T = 0$, they will lower the critical temperature but 
the zero temperature gap will be much less affected. This leads naturally 
to an increase of the $\Delta_{0} / T_{c}$ ratio. In order for the 
number of these scattering sources to be temperature dependent, we 
have merely to take them as bosons, corresponding to a proper kind of 
collective modes of our system. Although other kind of fluctuations or 
modes may be considered, the simplest and most natural choice is 
phonon scattering. As we will see the typical energy needed for these 
bosons is in reasonable agreement with the frequencies available for 
phonons in these compounds.

The idea of explaining a large value of $\Delta_{0} / T_{c}$ by a 
decrease of  $T_{c}$ is already present in the literature, but to our 
knowledge it has not been put to work specifically in the case of d-wave 
superconductors. It is actually the usual qualitative picture behind the 
enhanced value of  $\Delta_{0} / T_{c}$ in strongly coupled standard 
superconductors : the argument is that thermally activated phonons tend 
to destroy superconductivity and lower $T_{c}$ while the zero 
temperature gap is not so affected since there are no real phonons 
present at T = 0. In the context of high $T_{c}$ superconductors Lee 
and Read \cite{lr}, noticing the strong inelastic scattering 
experimentally observed, have already proposed qualitatively this kind 
of mechanism to suggest a lowered $T_{c}$. Here we rely specifically 
on the fact that the order parameter in high $T_{c}$ superconductors 
changes sign to obtain an important effect, compatible with experiment. 
We will more precisely assume the d-wave symmetry, as it is most 
often done, although our mechanism actually requires only basically 
that the order parameter takes different signs on the Fermi surface.

In order to explore this kind of explanation and see if it can work 
quantitatively for high $T_{c}$ compounds, we take the simplest 
possible model which retains all the qualitative features of our picture. 
Specifically we consider a class of models which has already been 
studied by Preosti, Kim and Muzikar \cite{paul} in the presence of 
impurity scattering : we mimic a d-wave superconductor by taking an 
order parameter which takes a constant value $\Delta_{+} $ on some 
parts of the Fermi surface and the opposite value $\Delta_{-} = -
\Delta_{+} $ on the rest of the Fermi surface. We immediately 
specialize to the situation where the $+$ and $ -$ regions have equal 
weight, as it is the case when they are related by symmetry. We assume 
that bosons scatter electrons from the $+$ to the $ -$ region and vice-
versa, and for simplicity we retain only those bosons. We take a simple 
Einstein spectrum with frequency $\Omega $ for these bosons, with 
coupling constant $\lambda $ to the electrons. We assume the pairing 
interaction to have a characteristic energy much higher than $T_{c}$ 
and $ \Omega $, and take a weak coupling description for the pairing. 
Therefore we do not make any specific assumption on the pairing 
mechanism : it may originate from pure Coulomb interaction or from 
spin fluctuations, or even have a more intricate origin. Again for 
maximum simplicity we keep only a constant repulsive pairing 
interaction between the $+$ and the $ -$ regions.
We do not expect any considerable quantitative changes from all these 
simplifications, all the more since it is known that the  $\Delta_{0} / 
T_{c}$ ratio is quite robust.

With all these simplications the Eliashberg equations at temperature $T$ 
read for our model :
\begin{eqnarray}
{\Delta }_{\pm ,n}{Z}_{\pm ,n}=\pi T \sum _{m} {\Lambda }_{n-m} 
\frac{{\Delta }_{\mp,m}}{{({\omega }_{m}^{2}+{\Delta 
}_{\mp,m}^{2})}^{1/2}}
\label{eq1}
\end{eqnarray}
\begin{eqnarray}
{\omega }_{n}({Z}_{\pm ,n}-1)=\pi T \sum _{m} {\lambda }_{n-m} 
\frac{{\omega }_{m}}{{({\omega }_{m}^{2}+{\Delta 
}_{\mp,m}^{2})}^{1/2}}
\label{eq2}
\end{eqnarray}
Here ${\Delta }_{\pm ,n}$ and ${Z}_{\pm ,n}$ are the order 
parameter and the renormalization function at the Matsubara frequency 
$\omega_{n} =   \pi T (2n+1)$ in the $\pm $ regions. The effective 
frequency-dependent interaction ${\Lambda }_{n } = {\lambda }_{n } 
- k $ contains the pairing interaction  $k$ , with a cut-off frequency 
$\omega_{c}$ and the boson mediated interaction ${\lambda }_{p} = 
\lambda \Omega^{2}/ (\Omega^{2} + \omega_{p}^{2}) $ with 
$\omega_{p} = 2 \pi pT$ the boson Matsubara frequency. As 
mentionned above we have $\omega_{c} \gg \Omega$  and $T_{c}$ . 
When we specialize to d-wave symmetry and insert the corresponding 
relation $\Delta_{-,n} = -\Delta_{+,n} \equiv \Delta_{n} $ into these 
equations, we obtain  $ Z_{-,n} = Z_{+,n} \equiv Z_{n} $ and find 
that $ \Delta_{n}$ and $ Z_{n} $ satisfy Eq.(1) and Eq.(2) (with $ 
\Delta_{\pm,n}$ and $ Z_{\pm,n} $ replaced by $ \Delta_{n}$ and $ 
Z_{n} $ ) provided that we change the sign in front of ${\Lambda }_{n 
}$. The resulting equations are just the ones obtained in standard strong 
coupling theory, except that the roles are reversed between the boson 
and the Coulomb terms : the boson term is repulsive and the Coulomb 
one attractive.

We have solved these equations directly both for the change of critical 
temperature and for the zero temperature gap. However it turns out to 
be much more convenient to eliminate the pairing interaction and the 
cut-off in favor of the critical temperature $T^{0}_{c}$ in the absence 
of boson scattering. This is done by taking explicitely into account that, 
for $ \Omega \ll \omega_{n} \ll \omega_{c}$ , $\Delta_{n} $ goes to a 
constant and  $ Z_{n} $ goes to 1. Let us first consider the calculation 
of $T_{c}$ , where $ \Delta_{n}$ gets very small and $ Z_{n} $ takes 
just its normal state value. We call $\Delta_{\infty} $ this large 
frequency limit of $\Delta_{n} $ and set $ \Delta _{n}{Z}_{n} = 
\Delta_{\infty} +  d_{n}$. Since, as can be checked, the pairing term 
dominates in this range we obtain from Eq.(1) and (2) (after taking into 
account that $ \Delta_{n}$ and $ Z_{n} $ are even functions of  
$\omega_{n}$) that $\Delta_{\infty}$ satisfies :
\begin{eqnarray}
{\Delta }_{\infty }= 2 k \pi T \sum _{m=0}^{\omega_{c}} 
\frac{{\Delta }_{\infty} + d_{m}} {{\omega }_{m}{Z}_{m}}
\label{eq3}
\end{eqnarray}
This leads to deal with  $ S = \pi T \sum  d_{m}/ {\omega 
}_{m}{Z}_{m}$ where the upper boundary can be taken as infinity 
since the sum converges. We have also to consider $ S' = \pi T \sum 1 / 
{\omega }_{m}{Z}_{m}$ where we have to keep the cut-off, but this 
can be expressed in terms of $T^{0}_{c}$ as $ S' = S_{Z} + (1/2) \ln 
(T^{0}_{c}/T) + 1/2k $ with $ S_{Z} = \pi T \sum (1/{Z}_{m}-1) / 
{\omega }_{m}$ where again  infinity can be taken as upper boundary. 
This leads to ${\Delta }_{\infty }= - S / [ S_{Z} + (1/2) \ln 
(T^{0}_{c}/T)] $. When this is carried into Eq.(1) this gives the 
numerically convenient eigenvalue problem :
\begin{eqnarray}
d_{n}= - \pi T \sum _{m=0}^{\infty} ({\lambda }_{n-m} + {\lambda 
}_{n+m+1})  \
\frac{{\Delta }_{\infty}+ d _{m}} {{\omega }_{m}{Z}_{m}}
\label{eq4}
\end{eqnarray}
which is satisfied when $T = T_{c}$ . Note that a similar procedure 
could be applied to the standard Eliashberg equations.

We are left with only two parameters, namely the reduced frequency  $ 
\Omega / T^{0}_{c}$ and the coupling strength $ \lambda $ of the 
bosons. We have plotted in Fig.1 the ratio $T_{c}/ T^{0}_{c}$, of the 
critical temperature $T_{c}$ compared to its value without coupling 
$T^{0}_{c}$, for various values of the ratio $ \Omega / T^{0}_{c}$ 
going from 0.2 to 1. Naturally $T_{c}$ decreases with increasing $ 
\lambda $ since boson scattering is pair breaking. High values of $ 
\Omega / T^{0}_{c}$ are not of much interest for us since they 
correspond to a full weak coupling regime, and the ratio $\Delta / 
T_{c}$ will merely be given by the standard BCS value. Similarly large 
values of $ \lambda $ lead to a strong decrease of  $T_{c}$ as can be 
seen in Fig.1 . This produces a large  $ \Omega / T_{c}$ and leads 
again to the BCS value for  $\Delta / T_{c}$. On the other hand the 
result in the low frequency limit  $ \Omega / T^{0}_{c} \rightarrow  0 
$ is easily obtained. Indeed in this case ${\omega }_{n}{Z}_{n} = \pi  
T (2n + 1 + \lambda ) $ and $ {\lambda }_{n-m} + {\lambda 
}_{n+m+1} = \lambda  \delta _{n,m} $. This leads from Eq.(4) to an 
explicit expression for the $n$ dependence of $d_{n}$ and to the result 
$ \ln (T^{0}_{c}/ T_{c}) = \psi ( \lambda +1/2) - \psi (1/2) $ where $ 
\psi $ is the digamma function. This is just an Abrikosov-Gorkov result 
\cite{ag}, as might have been anticipated since bosons behave as 
impurities in the $ \Omega \rightarrow  0 $ limit \cite{br}. This 
analytical result is actually in very good agreement with our numerical 
calculations for $ \Omega = 0.2 $. A noticeable feature of Fig.1 is the 
strong sensitivity of $T_{c}$ on boson scattering even for moderate 
coupling strength. Indeed it is severely reduced already for $ \lambda \ll 
1$. For example the slower decrease of $T_{c}$ corresponds to the 
case  $ \Omega / T^{0}_{c} \rightarrow  0 $, but even in this case the 
slope for small $ \lambda $ is $ - \pi ^{2}/2 $.

\begin{figure}
\vbox to 7cm{\hspace{-3mm} \epsfysize=6.8cm \epsfbox{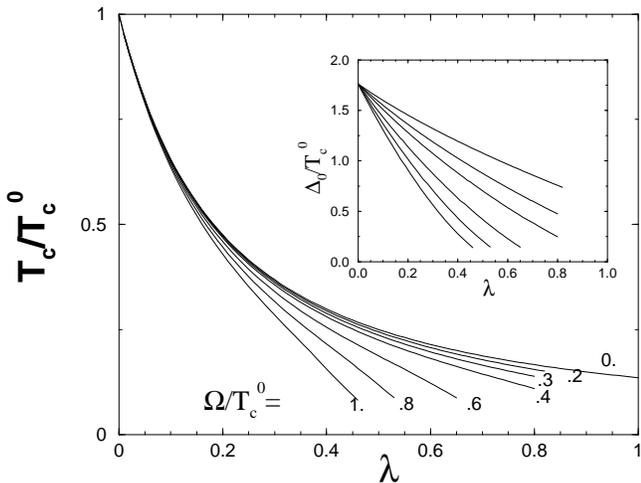} }

\caption{$T_{c}/T_{c}^{0}$ as a function of the coupling strength $\lambda$
for $\Omega/T_{c}^{0} = 0., 0.2, 0.3, 0.4, 0.6, 0.8 $ and $1 .$
Insert: $\Delta_{0}/T_{c}^{0}$ as a function of $\lambda$, for the 
same values of $\Omega/T_{c}^{0}$ ( $\Omega=0$ omitted).}

\label{figure1}
\end{figure}

Let us consider now the calculation of the zero temperature gap, where 
the situation is somewhat more complicated. In the $ T \rightarrow 0 $ 
limit  $ \Delta_{n}$ and $ Z_{n} $ become functions $ \Delta( \omega ) 
$ and $ Z( \omega ) $ of the continuous variable $ \omega _{n} \equiv 
\omega $. As for the calculation of $T_{c}$ we set $ \Delta ( \omega ) 
Z( \omega ) = \Delta_{\infty} +  d( \omega ) $. In the same way the 
pairing term dominates in Eq.(1) for large $ \omega $ which leads to the 
following equation for $  \Delta_{\infty}$ :
\begin{eqnarray}
\Delta_{\infty} = k \int _{0}^{\omega _{c}} d \omega ' \frac{\Delta ( 
\omega ' )}{[{ \omega '} ^{2} + \Delta ^{2} ( \omega ' )] ^{1/2}}
\label{eq5}
\end{eqnarray}
where naturally  $ \Delta ( \omega ) $ in the right-hand side is expressed 
in terms of $ Z( \omega )$ and $ d(\omega )$ . For $ d(\omega )$ we 
are left with :
\begin{eqnarray}
d(\omega ) = - \frac{1}{2} \int _{- \infty}^{\infty} d \omega ' \  
\lambda _{\omega - \omega '} \ \frac{\Delta ( \omega ' )}{[{ \omega '} 
^{2} + \Delta ^{2} ( \omega ' )] ^{1/2}}
\label{eq6}
\end{eqnarray}
with $ \lambda _{ \omega} = \lambda \Omega^{2}/ (\Omega^{2} + 
\omega^{2})$ and :
\begin{eqnarray}
\omega ( Z( \omega )-1 ) = \frac{1}{2} \int _{- \infty}^{\infty} d 
\omega ' \  \lambda _{\omega - \omega '} \ \frac{ \omega ' }{[{ \omega 
'} ^{2} + \Delta ^{2} ( \omega ' )] ^{1/2}}
\label{eq7}
\end{eqnarray}
where we can naturally use the even parity of $ \Delta( \omega ) $ and $ 
Z( \omega ) $. As we have done above for $T_{c}$ we can use the 
weak coupling expression for the zero temperature gap $ \Delta 
_{BCS} = 1.76 \ T^{0}_{c} $ for $ \lambda = 0 $ , obtained from 
Eq.(5) by setting $ \Delta( \omega ) = \Delta_{\infty} = \Delta _{BCS} 
$, to eliminate the cut-off $\omega _{c}$ and $k$. This leads to : 
\begin{eqnarray}
\ln  \frac{\Delta_{\infty}}{ \Delta _{BCS}} = \! \int _{0}^{\infty } \! \!
d \omega  \ \frac{\Delta ( \omega ) / \Delta_{\infty}}{[{ \omega } ^{2}
+ \Delta ^{2} ( \omega )] ^{1/2}} - \frac{1}{[{ \omega } ^{2} 
+\Delta_{\infty} ^{2}] ^{1/2}}
\label{eq8}
\end{eqnarray}
This last equation does not provide an explicit expression for 
$\Delta_{\infty}$ in contrast with what we have for $T_{c}$. However 
this is not in practice a problem, since it can be easily included in 
the iteration procedure used to solve numerically Eq.(6),(7) and (8). 
In order to find the gap  $ \Delta _{0}$ we still have to continue $ \Delta 
( \omega )$ and $ Z( \omega )$ analytically toward the real frequency 
axis into $\bar{\Delta} ( \nu ) \equiv  \Delta ( - \dot{\imath} \nu ) $ and 
$\bar{Z} ( \nu ) \equiv  Z ( -  \dot{\imath} \nu ) $, and solve $ 
\bar{\Delta} ( \Delta _{0}) = \Delta _{0}$. This is done by using the 
explicit expression for this continuation \cite{msc}. Note that this 
analytic continuation lowers noticeably the gap values, compared to the 
naive evaluation $  \Delta _{0}= \Delta (0)$.

Just as for $T_{c}$ , the low boson frequency limit $ \Omega 
\rightarrow 0 $ is of particular interest. Indeed it leads to a very simple 
model of strongly interacting fermions with quite non trivial results. 
Naturally we have in this limit to let $ \lambda $ increase in such a way 
that $ \lambda \Omega $ stays finite otherwise one obtains trivially the 
BCS result, as it is clear from the equations found below. For this case 
Eq.(6) and (7) lead to algebraic equations because the lorentzian coming 
in the integrals gets very narrow. One obtains $ Z( \omega ) =  1 +  \pi 
\lambda \Omega /2 [{ \omega } ^{2} + \Delta ^{2} ( \omega )] ^{1/2}$ 
and $ d( \omega ) = -  \pi \lambda \Omega \Delta ( \omega ) /2 [{ \omega 
} ^{2} + \Delta ^{2} ( \omega )] ^{1/2}$ , giving for $ \Delta ( \omega 
)$ the simple equation :
\begin{eqnarray}
\Delta ( \omega ) = \Delta_{\infty} -  \pi \lambda \Omega \frac{ \Delta ( 
\omega )} {\sqrt{{ \omega } ^{2} + \Delta ^{2} ( \omega )} }
\label{eq10}
\end{eqnarray}
This equation (a fourth order equation) can be solved analytically and is 
very simple to solve numerically. The analytical continuation is merely 
obtained by replacing $ \omega ^{2} $ by  $ - \nu ^{2} $ in Eq.(9). 
However one can see easily that the equation $\bar{\Delta} ( \nu ) /  
\Delta_{\infty} = 1 - K \bar{\Delta} ( \nu ) / [ -{\nu } ^{2} + 
\bar{\Delta} ^{2} ( \nu )] ^{1/2} $, where $ K = \pi \lambda \Omega /  
\Delta_{\infty}$,  has no purely real solution when $ \nu /  
\Delta_{\infty}$ becomes larger than $ \nu  _{0} /  \Delta_{\infty} = ( 1 
- K ^{2/3}) ^{3/2}$ , corresponding to a gap $ \Delta _{0} = 
\bar{\Delta} ( \nu _{0}) =  \Delta_{\infty} ( 1 - K ^{2/3}) $. Indeed, 
beyond this point, $\bar{\Delta} ( \nu )$ gets complex and the density 
of states is no longer zero. It is quite interesting to note that, in this 
limit, this nonzero density of states occurs before the equality $ 
\bar{\Delta} ( \Delta _{0}) = \Delta _{0}$ is reached. To be complete 
we have to find the value of $  \Delta_{\infty}$ in this limiting situation. 
The integral in the defining Eq.(8) can actually be performed 
analytically (essentially by taking $ \Delta ( \omega)$ as the variable) 
and is merely equal to $ \pi K / 4 $. Then $  \Delta_{\infty}$ is obtained 
as the solution of the simple transcendental equation $ \Delta_{\infty} / 
\Delta _{BCS} = \exp ( - \pi ^{2} \lambda \Omega / 4 \Delta_{\infty})$ 
. Note that this equation has always a single solution in the physical 
range $ K < 1 $ . 

Coming back to our problem, we have, for $ \Omega  / T^{0}_{c}= 
0.2$ and a varying coupling strength $ \lambda  $, compared the result 
obtained for $ \Delta_{\infty}$ in the limiting situation we have just 
considered with the general calculation we have performed for any $ 
\Omega / T^{0}_{c}$. The agreement is quite good. From this it would 
be tempting to conclude that, for small $ \Omega / T^{0}_{c}$, we can 
also conveniently extract the gap itself from this analytical solution. 
This is unfortunately not true : a good agreement for imaginary 
frequencies does not imply that the analytical continuations to the real 
frequency axis agree quite closely, since this analytical continuation is 
very sensitive to small differences as it is well known. And indeed in 
our case there is a sizeable difference between the gaps obtained by the 
two methods. The results of our calculations for the zero temperature 
gap are displayed in the insert of Fig.1. They show that, even for 
small $ \lambda 
$,  $ \Delta _{0}$ is quite sensitive to boson scattering, although the 
effect is not as strong as for $T_{c}$. We note in particular that the 
simple expectation that the gap would be essentially unchanged at $T = 
0$ because no bosons are present is not correct. 

\begin{figure}
\vbox to 6cm{\hspace{-4mm} \epsfxsize=8.5cm \epsfbox{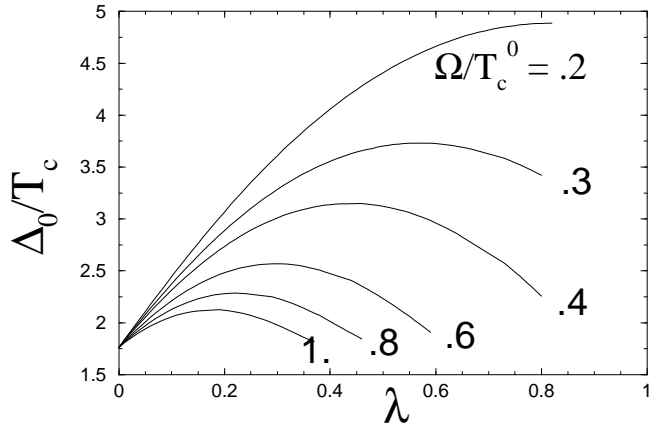}}

\caption{ Ratio of the gap over the critical temperature
$\Delta_0/T_{c}$ as a function of the coupling strength
$\lambda$, for fixed values of $\Omega/T_{c}^{0}= 0.2, 0.3, 0.4, 0.6,0.8,1.$}

\label{figure2}
\end{figure}

Finally our results for the ratio $ \Delta _{0} / T_{c}$ are shown in 
Fig.2  for the interesting range of values for the parameter $  \Omega  / 
T^{0}_{c}$. We note first that, for  $  \Omega  / T^{0}_{c} = 1 $, the 
result does not depart much from the BCS result. Naturally this is even 
more so for higher values of $  \Omega  / T^{0}_{c} $ for which the 
results are not displayed. In the same way we find as expected that, for 
large values of $ \lambda $, $ \Delta _{0} / T_{c}$ decreases toward 
the BCS value. Because this is of little interest for our purpose, we 
have not explored further this regime where numerical calculations get 
more difficult since one has to deal with very different energy scales. 
The most interesting feature of our results is naturally the maximum 
obtained for $ \Delta _{0} / T_{c}$ at intermediate coupling strength. 
This maximum increases with decreasing $  \Omega  / T^{0}_{c}$ 
while the $ \lambda $ corresponding to the maximum increases at the 
same time. In particular for $  \Omega  / T^{0}_{c}= 0.2 $ we find $ 
\Delta _{0} / T_{c}$ close to 5. Clearly this trend continues as $ 
\Omega $ goes to zero. Indeed as it is clear from above, the gap is 
independent of $ \lambda $ in this limit whereas we have seen that  
$T_{c}$ decreases toward zero. Therefore we can in principle obtain a 
ratio  $ \Delta _{0} / T_{c}$ as high as we like. However this would 
correspond to extreme parameter values. 

On the other hand we find among our results a range which is quite 
compatible with experiments. Indeed for $  \Omega  / T^{0}_{c}= 0.4 
$ we obtain a broad maximum for $ \lambda \approx .4 $ with  $ \Delta 
_{0} / T_{c} \approx 3.2 $. This fairly small value of $ \lambda $ is 
quite reasonable. The value of $ \Delta _{0} / T_{c}$ is already quite 
consistent with experimental data, all the more if we take into account 
that anisotropy of the order parameter is likely to raise $ \Delta _{0} / 
T_{c}$ by itself, as it does for d-wave in weak coupling where this 
ratio is raised by 20 \%. To be more specific, for $  \Omega  / 
T^{0}_{c}= 0.4 $, we find  $ \Delta _{0} / T_{c} \geq 3.1 $ for $ 
0.36 \leq \lambda \leq 0.52 $ . For a typical value of $ T_{c}$ = 90 K , 
we find that the range of boson frequency goes from 115 K to 170 K. 
For $  \Omega  / T^{0}_{c}= 0.3 $ we obtain correspondingly $ \Delta 
_{0} / T_{c} \geq 3.6 $ for $ 0.42 \leq \lambda \leq 0.73 $, with $  
\Omega $ ranging between 100 K to 170 K. This is a frequency range 
where an important weight for phonons is known to exist in these 
compounds. Therefore, at least for optimally doped or overdoped 
compounds, our explanation for the high value of $ \Delta _{0} / 
T_{c}$ is completely coherent with experiment, which is quite 
satisfactory. For markedly underdoped compounds the general situation 
is no so clear and it is likely that the very high values observed in this 
case require an additional physical source which might for example be 
disorder. 

However a very striking feature of our interpretation is that it requires a 
fairly high value of $ T^{0}_{c}$, that is the critical temperature 
without bosons, ranging from 290 K to 420 K for  $  \Omega  / 
T^{0}_{c}= 0.4 $ and from 330 K to 560 K for  $  \Omega  / 
T^{0}_{c}= 0.3 $. Naturally it would be quite desirable to check 
experimentally this physical aspect of our model. One possible way 
would be to send a flux of phonons with the proper frequency to see if 
$ T_{c}$ is affected as expected (these phonons could be generated 
themselves by tunnel junctions). Another much more interesting, 
though speculative, possibility is to try to raise $ T_{c}$ toward $ 
T_{c} ^{0}$ under static conditions. This could be done through a 
shift of the phonon spectrum, for example by applying high pressure in 
order to lower the number of phonons which participate in the decrease 
of $ T_{c}$. Actually it is known that, for Hg compounds, $ T_{c}$ 
increases with pressure, which is compatible with our model. A test 
would be to measure also $ \Delta _{0}$ under pressure and check that 
it is less sensitive to pressure than  $ T_{c}$. This would be a clear 
indication that one can hope to raise  $ T_{c}$ even further by 
modification of the phonon spectrum. It is also important to keep in 
mind that it is often possible to obtain effectively an increase of pressure 
by proper chemical substitution in the compound. Therefore an 
understanding of the effect of pressure would open the way to a 
possible chemical increase of  $ T_{c}$. Finally it is tempting to believe 
that the pseudogap observed above $ T_{c}$ in underdoped 
compounds might be related to $ T^{0}_{c}$ and could be obtained by 
treating our model beyond mean-field theory. 

In conclusion we have shown that the adverse effect on d-wave 
superconductors of boson scattering between regions with opposite 
sign of the order parameter provide a simple and natural explanation for 
the high values of  $ \Delta _{0} / T_{c}$ observed experimentally.

We are very grateful to A. A. Abrikosov and D. Rainer for very 
stimulating discussions.

* Laboratoire associ\'e au Centre National de la Recherche Scientifique 
et aux Universit\'es Paris 6 et Paris 7.

\end{multicols}
\end{document}